# 基于卫星遥感资料的近海海上通信环境研究*

倪晗玥[1,2]，杨劲松[1,2**]，任林[2]，李晓辉[2]，董昌明[3]，陈文[1]

（1.上海交通大学，上海 200240；
2.自然资源部第二海洋研究所，浙江 杭州 310012；
3.南京信息工程大学，江苏 南京 210044）

【摘　要】　海气界面通量等要素会影响海上通信的可靠性和效率。相较于分布零星的海上现场观测数据，卫星遥感数据有覆盖范围广、时间跨度长的优势，基于星载合成孔径雷达风速数据和再分析数据以及浮标实测数据，结合神经网络方法，利用COARE V3.5算法计算海气界面的动量通量、感热通量和潜热通量。研究发现，经过神经网络校正后的SAR风速数据在计算海气通量时，与浮标实测风速的一致性得到了显著提升，摩擦风速的偏差从-0.03 m/s降低到0.01 m/s，风应力的偏差从-0.03 N/m²降低到0.00 N/m²，拖曳系数的偏差从-0.29降低到-0.21，潜热通量的偏差从-8.32 W/m²降低到5.41 W/m²，感热通量的偏差从0.67 W/m²减小到0.06 W/m²。研究结果表明，经过神经网络校正后的SAR风速资料能够提供更加可靠的海上通信环境数据。

# Research on the Offshore Marine Communication Environment Based on Satellite Remote Sensing Data

NI Hanyue[1,2], YANG Jingsong[1,2], REN Lin[2], LI Xiaohui[2], DONG Changming[3], CHEN Wen[1]

(1. Shanghai Jiao Tong University, Shanghai 200240, China;
2. Second Institute of Oceanography, Hangzhou 310012, China;
3. Nanjing University of Information Science and Technology, Nanjing 210044, China)

[Abstract]　Air-sea interface fluxes significantly impact the reliability and efficiency of maritime communication. Compared to sparse in-situ ocean observations, satellite remote sensing data offers broader coverage and extended temporal span. This study utilizes COARE V3.5 algorithm to calculate momentum flux, sensible heat flux, and latent heat flux at the air-sea interface, based on satellite synthetic aperture radar (SAR) wind speed data, reanalysis data, and buoy measurements, combined with neural network methods. Findings indicate that SAR wind speed data corrected via neural networks show improved consistency with buoy-measured wind speeds in flux calculations. Specifically, the bias in friction velocity decreased from -0.03 m/s to 0.01 m/s, wind stress bias from -0.03 N/m² to 0.00 N/m², drag coefficient bias from -0.29 to -0.21, latent heat flux bias from -8.32 W/m² to 5.41 W/m², and sensible heat flux bias from 0.67 W/m² to 0.06 W/m². Results suggest that the neural network-corrected SAR wind speed data can provide more reliable environmental data for maritime communication.

[Keywords]　satellite remote sensing; air-sea flux; marine communication environment; neural network

收稿日期：2024-09-23
*基金项目：国家自然科学基金“近岸条件下北斗导航反射信号海面风速反演能力增强机制研究”（42306200）；国家自然科学基金–青年科学基金项目“基于深度学习的热带气旋风场重构及风暴潮模拟”（42306216）；国家重点研发计划项目“环境遥感巡查与立体智能感知”（2022YFC3103101）
**通信作者

## 0 引言

近些年来人类海上活动日渐频繁，海上旅游业务、近海水产养殖和海上资源勘探等经济活动发展迅速，海上通信的需求也日渐增大。目前海上通信系统主要被分为几个部分，分别是基于空域的海上通信，主要以海上卫星为主；基于陆地的海上通信，主要由陆地基岸的海上通信构成，是地面蜂窝的拓展；以及基于海域的海上通信，是由主要包括岛屿、船舶、海上航空器和无人机等组成的无线通信网络。与陆地不同，海洋环境对海上通信的影响更加复杂多样、海上基站部署困难、通信条件恶劣 [1-2]。海水温度和海上大气温度的差异会影响海上大气的稳定度，导致海上大气分布不均；海气温差还会产生水蒸气，改变海面上方湿度；同时海上的天气状况，雨、雪雾等也会对无线电波信号造成衰落，进而影响接收场强和通信质量 [3-4]。这些复杂多变的海上状况会影响海上电磁波的传播，使得海上的无线信道特征明显不同于陆地，因此需要根据具体的海况建立适用于海域通信的海上无线信道模型，来满足不同的海上通信需求 [5-7]。Qiu 等 [8] 针对特殊海上环境中的无线网络通信，讨论了在含盐雾介质中电磁波的衰减系数，通过将含盐雾的介电常数替换为氯化钠溶液的介电常数，并引入视线考虑的电磁衰落模型。最后，设计了一个海洋电磁波衰减模型，为海上建立无线通信网络提供建议。张颖 [9] 等为提高对海上稀疏信道的估计性能，去除海上通信信道传输时外界环境因素的干扰，提出了一种基于奇异值分解优化观测矩阵的快速贝叶斯匹配追踪稀疏信道估计优化算法。该方法在建模时考虑了信号传播时，通信功率受大气吸收电波和海面反射等因素影响而存在的损耗，计算机仿真实验表明，与传统的信道估计算法相比，该算法能够提高信道估计的精确度。

由于海上环境对海上通信的复杂影响以及海上通信面临的技术挑战，对其进行深入研究，对建立海上通信无线信道模型、发展跨域协同的海上通信系统等具有重要意义，对满足国家日益增长的海上通信需求尤为重要 [10-13]。其中卫星遥感数据凭借其观测区域范围大、观测时间跨度长、数据连续性好等优势，已成为研究海上环境的重要数据源。Crespo[14] 等在 2019 年用 CYGNSS 卫星系统风速数据获得的热通量产品与浮标实测的真实值进行对比，发现用 GYCNSS 测得的风速得到的感热通量和潜热通量与用浮标数据获得的通量只具有较好的一致性，GYCNSS 卫星系统可提供大范围长时间尺度且可靠性较高的热通量产品。Li[15] 等在 2022 年分析了 CYGNSS 热通量产品在赤道区域的数据精度，在试验时同时使用了 CYGNSS 二级热通量数据 1.0 版本和 1.1 版本，以及用 CYGNSS V3.1 风速数据和 MERRA-2 湿度数据计算得到热通量，分别与浮标数据进行对比分析后发现，风速数据的准确性对热通量产品的准确性具有较大影响，相较于 V1.0 热通量产品，用更高精度的风速数据得到的热通量的均方根误差减小了 17%。该研究还讨论了 CYGNSS 热通量产品在大西洋、太平洋和印度洋的准确性，并发现在大西洋的热通量产品表现更好。合成孔径雷达是研究近岸海面风速的主要遥感数据源之一，相较于微波散射计，合成孔径雷达的分辨率空间更高，可以提供更为精细的海面风速分布情况，已有研究表明，经过一定校正后的合成孔径雷达风速数据与浮标实测数据有较好的一致性，可用于对海上环境进一步的研究 [16]。

综上所述，海洋环境对海上通信有重要影响，在建立信道模型时去除海上信号传输受海上环境的影响后，质量更高的遥感数据能够提供更加准确的海上环境信息。利用卫星遥感数据能够为海上通信环境以及提高海上通讯质量的研究提供新思路。本文基于“哨兵一号”卫星双星系统（Sentinel-1 A/B）的合成孔径雷达（SAR, Synthetic Aperture Radar）遥感风速数据和 MERRA-2 再分析数据以及 NDBC 浮标数据，结合 BP 神经网络方法，利用 COARE V3.5 算法探讨了影响海上通信的海气界面要素。

## 1 数据与方法

### 1.1 研究数据

（1）Sentinael-1 SAR 风速数据

遥感风速数据源于美国国家海洋和大气管理局（NOAA, National Oceanic and Atmospheric Administration）下的环境信息中心（NCEI, National Centers for Environmental Information）提供的 2019 年 -2023 年的“哨兵一号”二级（Level-2）10 m 高风速数据产品（www.ncei.noaa.gov），其中哨兵 1B 于 2021 年底发生故障，停止传输数据。该产品的原始数据是 NOAA 从欧空局获取的，用 CMOD4 地球物理函数模式反演并用 ASCAT 散射计数据和 ECMWF（European Center for Medium-Range Weather Forecasting）以及 GFS（Global Forecast System）数值模式数据进行交叉验证和质量控制 [17]，该模型是目前较为完善的用风速反演地球物理模式函数，被广泛应用于 C 工作波段的微波遥感数据的反演。SAR 影像的空间覆盖范围为 180° W—16° W，时间分辨率不定，空间分辨率为 500 m。

（2）NDBC 浮标数据

国家资料浮标中心（NDBC, National Data Buoy Center）

是美国 NOAA 机构下设的专门对海洋资料浮标的进行研究与管理的专门机构，海洋锚系浮标常年在海上固定位置工作，用于观测海表面风场、海表面温度等，为海洋领域的研究提供了大量珍贵的现场实测数据。本文选用了分布在美国东西海岸的 36 个近岸 NDBC 浮标在 2019 年到 2023 年的观测数据，时间分辨率为 10 min，包括风速、气温、海温、气压、有效波高和有效波周期。NDBC 浮标高度普遍大概在海面 3 m 以上，风速仪的高度距海面约 3~5 m，由于所选用的 SAR 数据为海表 10 m 高风速，故需对浮标实测的风速进行转换。近海面层风速随高度增加呈对数分布，并受到大气稳定度和海面粗糙度的影响 [18-20]。在大气层结中性条件下，对数风廓线可简化 [21-22] 为：

$$\frac{U_{10}}{U_z}=\frac{\ln\left[\frac{10}{z_0}\right]}{\ln\left[\frac{z}{z_0}\right]} \tag{1}$$

式中，$U_{10}$ 和 $U_z$ 分别表示在 10 m 高度的风速和在 $z$ 高度的风速，$z$ 表示距海面的高度，$z_0$ 表示海面粗糙度。

基于研究区域相近且同为 NDBC 浮标，本文参照陈克海等 [23] 使用的转换公式进行转换，转换公式如下：

$$U_{10}=8.87403\times U_z/\ln\left(z/0.0016\right) \tag{2}$$

其中 $z$ 表示距海面的高度，$U_{10}$ 和 $U_z$ 分别表示 10 m 高风速和在 $z$ 高度的风速。

（3）MERRA-2 再分析数据

再分析数据是用于估算海洋表面和近表面热力学变量的重要数据，NASA 制作的全球大气再分析数据集 MERRA-2（Modern-Era Retrospective Analysis for Research and Applications, Version 2）（gmao.gsfc.nasa.gov/reanalysis/MERRA-2）使用数据同化技术，将所有可用的现场观测数据和卫星数据与全球大气模型提供的大气状态估计量结合了起来 [24]。MERRA-2 是用于估算感热通量和和潜热通量可靠的数据产品，具有相对较高的空间和时间分辨率。本研究使用的 MERRA-2 数据中 10 m 高的比湿数据 QV10M，时间分辨率为 1 小时，空间分辨率为 0.5° ×0.625° 的网格。

（4）CMORPH 降水数据

文献 [25-26] 表明，降水对遥感方法反演海面风速的精度有一定影响，为了提高对 SAR 风速数据的校正精度，需要将降水数据加入到校正过程中。CMORPH 降水资料（ncei.noaa.gov/products/climate-data-records/precipitation-cmorph）是多源卫星融合资料，该数据的质量在不同地域都有着较好的表现。本研究收集了 2019—2023 年的 CMORPH 降水数据，该数据从 NCEI 数据集网站下载，根据选取的 NDBC 浮标分布范围，选取了经度在 140° W—60° W，纬度在 20° N—50° N 范围的数据，空间分辨率为 8 km，时间分辨率为 30 min。

（5）数据预处理和偏差统计量

并非所有 SAR 风速遥感数据都与浮标位置在空间上重合，故在计算海气界面通量前，首先需要将不同数据源的数据在时间和空间上进行匹配，保证其时空同步性，建立匹配数据集。

参考 Xing 等 [27] 在对 HY-2A 散射计风速数据校正时采用的方法，将浮标风速和浮标风速的时间差控制在 5 min 内，空间匹配窗口为 250 m。针对 CMORPH 降水资料和 MERRA-2 比湿数据，在 SAR 风速资料与 NDBC 浮标资料生成的匹配数据集的基础上，用 15 min 和 4 km 以及 30 min 和 25 km 作为标准再次进行匹配，最终得到的匹配数据集共有 3 066 组匹配数据。

本次研究需评估校正前后的 SAR 风速资料以及 NDBC 浮标观测风速计算得到的海气界面通量之间的偏差，本文计算了三个偏差统计量，分别是平均偏差（bias）、均方根误差（RMSE, Root Mean Squared Error）和标准差（STD, Standard Deviation）[28]。

$$bias=\frac{1}{N}\sum_{i=1}^{N}(wspd_SAR_i-wspd_NDBC_i) \tag{3}$$

$$RMSE=\sqrt{\frac{1}{N}\sum_{i=1}^{N}(wspd_SAR_i-wspd_NDBC_i)^2} \tag{4}$$

$$STD=\sqrt{\frac{1}{N}\sum_{i=1}^{N}(wspd_SAR_i-wspd_NDBC_i-bias)^2} \tag{5}$$

上式中 $wspd_SAR_i$ 表示数据集中第 $i$ 个 SAR 风速数据，$wspd_NDBC_i$ 表示数据集中第 $i$ 个浮标数据，$N$ 为数据集的数据量。

### 1.2 研究方法

（1）COARE V3.5

目前国际上较先进的用来计算海气表面通量方法是 COARE（Coupled-Ocean Atmosphere Response Experiment）算法 [29]，该算法是基于 Monin-Obukhov（MO）相似理论，根据 1992 年 11 月 ~1993 年 2 月在西太平洋暖池进行了四个月的热带海洋与全球大气研究计划“耦合海气响应试验”（TOGA COARE, Tropical Ocean Global Atmosphere program and Coupled Ocean- Atmosphere Response Experiment）的分析资料，在 Liu[21] 等于 1979 年提出的算法上进行了改进。Monin-Obukhov 理论在假设大气边界层的常通量层是均匀稳定的基础上，认为湍动能主要由剪切和浮力生成，提出：

$$U_z-U_0=\frac{u_*}{k}(\ln\frac{z}{z_0}-\Psi) \tag{6}$$

其中$u_*$为磨擦速度，且$u_* = \sqrt{\tau / \rho}$; $z_0$为空气动力学粗糙度，它表示的是风速廓线外延到平均速度等于零那点所对应的高度。Ψ 为表示大气层结效应的函数，是 Obukhov 长度 $L$ 的函数。Geermaert[30] 给出了其通用形式：

$$\Psi = \begin{cases} 2\ln\left(\frac{1+\Phi_m^{-1}}{2}\right)+\ln\left(\frac{1+\Phi_m^{-2}}{2}\right)-2\tan^{-1}\left(\Phi_m^{-1}\right)+\frac{\pi}{2}, & z/L<0 \\ -5z/L, & z/L>0 \end{cases}$$
$$\Phi_m = \left(1-16z/L\right)^{-1/4}$$
$$L = -\frac{u_*^3 T_v}{kg\overline{w'T'_v}} \tag{7}$$

其中 $T_v$ 为虚温。

为了拓展该算法在中高风速条件下的适用性，Fairall[31] 等在 2003 年提出了 COARE V3.0 算法，引入 Taylor[32] 等给出的 Taylor01 和 Oost[33] 等提出的 OO02 两种参数化方案，这两种方案在考虑了真实海浪状态的前提下对海面空气动力学粗糙度进行计算。在有效波高和谱峰周期缺省时，COARE V3.0 算法能根据 Taylor01 方案利用海面 10 m 风速 $U_{10}$ 来对有效波高和谱峰周期进行参数化 [34]。Edson[31] 等人于 2013 年，对 COARE V3.0 版本进行了改进，提出了 COARE V3.5。COARE V3.5 版本在多个关键方面对前一版本的算法进行了改进，以更准确地模拟和预测海洋与大气之间的动量交换。与COARE V3.0相比，3.5 版本引入了对 Charnock 系数的重新参数化，这是一个表征海面粗糙度的关键参数，直接影响到风速与海面之间动量交换的计算。在 3.5 版本中，Charnock 系数的计算公式被修改为 $\alpha=mU_{10N}+b$，其中 $m$ 值为 0.001 7 $m^2/s$，$b$ 值为 -0.005，这一改变显著提高了在较高风速（超过 13 m/s）时对海气界面动量通量估算的精度。除了对 Charnock 系数的调整，COARE V3.5 还扩展了算法适用的风速范围，基于大量的现场观测实验（包括 RASEX、MBL、CBLAST 和 CLIMODE 项目），这些现场观测实验包含了不同的海况和大气稳定性条件，使算法能够更准确地处理高达 25 m/s 的风速情况。

在计算海气界面通量时，输入 NDBC 浮标测得的风速、气温、海温等，以及 MERRA-2 再分析数据的 10 m 高比湿，进行迭代得到各种近地面的特征尺度值，包括摩擦风速、拖曳系数、风应力、感热通量和潜热通量。

（2）BP 神经网络

BP 神经网络（BPNN, Back Propagation Neural Network）又称为前馈反向传播神经网络，是应用最为广泛的神经网络算法之一。在非线性多层网络中采用梯度下降算法，用反向传播误差来调节各神经元的权重。BP 神经网络由输入层、输出层和隐含层三个部分组成，当信号被传入输入层后，被隐含层神经元的作用函数计算后，信号被传播到输出神经元；将实际输出值与目标输出信号进行比较，若输出层没有得到预期的结果，则通过计算误差并将误差反向传播，不断调整权值和阈值，使得模型不断逼近期望误差，找到输入样本数据和输出样本数据之间最佳的映射关系 [35]。

在 BP 神经网络中，每个神经元通过激活函数来完成输入量到输出量之间的非线性映射，本文使用了 ReLu 激活函数：

$$f(x) = \max(x,0) \tag{8}$$

式中 $x$ 为输入量。

本研究使用了 PyTorch 机器学习库来实现神经网络的搭建与运行，搭建了三层 BP 神经网络结构，包括输入层 SAR 反演风速、大气稳定度和 CMORPH 降水数据 3 个神经元，输出层 NDBC 浮标观测风速 1 个神经元。初始模型训练集与测试集的划分比例为 60% 与 40%，初始训练批次为 1 500 次，隐含层神经元数量为 100，学习率为 0.005。其中大气稳定度由浮标测得的气温和海温确定，若海温大于气温，海气界面不稳定，大气稳定度为设为 0；若海温等于气温，海气界面中性稳定，大气稳定度设为 1；若海温小于气温，海气界面稳定，大气稳定度设为 2。

（3）技术路线

本文的技术路线如图 1 所示，首先将全数据集的 MERRA-2 比湿数据、气温、海温等数据分别和 SAR 风速、浮标风速输入 COARE V3.5 算法，计算出基于 SAR 风速资料和浮标实测风速资料的海气界面通量，包括摩擦风速、拖曳系数、风应力、潜热通量和感热通量。接着将全数据集分为 60% 的训练集和 40% 的测试集，用训练集数据对 BP 神经网络进行训练，用得到的模型对测试集 SAR 风速进行校正，得到更加接近浮标真实值的风速，再用 COARE V3.5 算法对结果校正的 SAR 风速进行计算得到海气通量，与浮标风速得到的进行对比，探究用 SAR 风速数据来研究海上通信环境的可行性。

## 2 结果分析

### 2.1 全数据集SAR遥感数据与实测数据对比

将全数据集的 3 066 组 SAR 风速和 NDBC 浮标风速进行对比分析，SAR 风速和浮标风速的密度散点图如图 2 所示，从图中可看出，大部分散点位于对角线的下方，散点的密度重心向 NDBC 浮标方向倾斜，在 5~15 m/s 风速区间内，SAR 风速要略低于 NDBC 浮标风速。

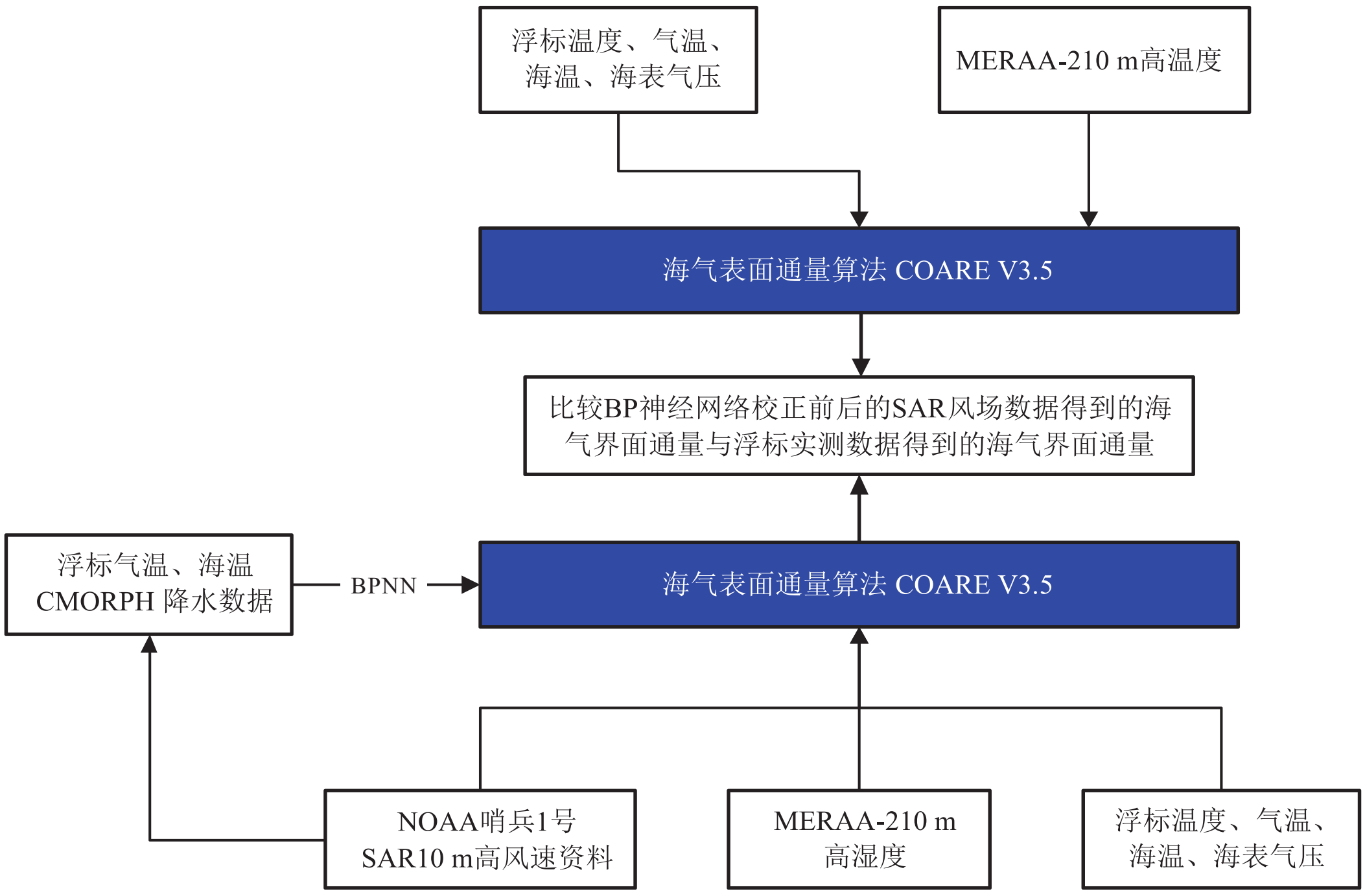

图1　技术路线图

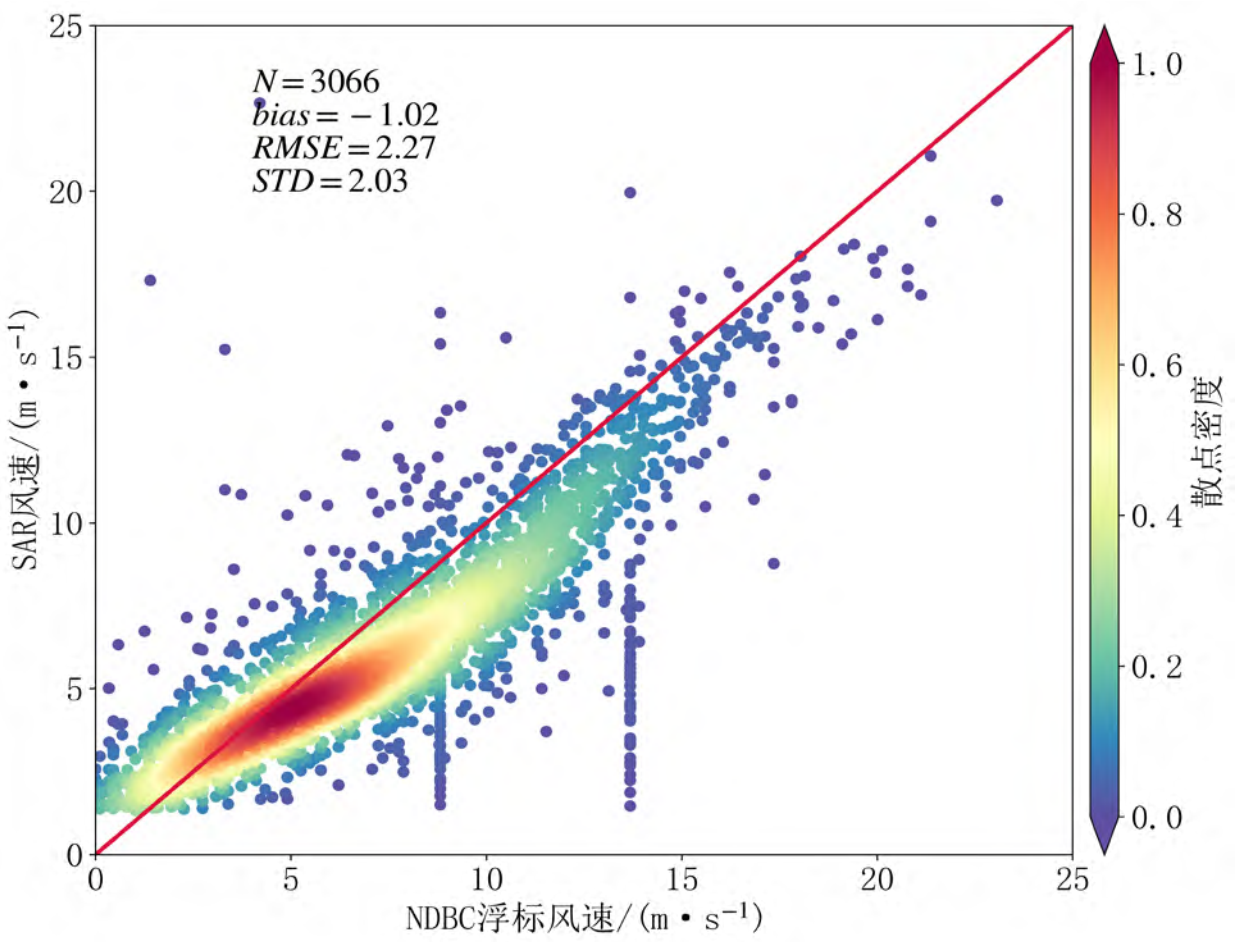

图2　全数据集SAR风速与浮标风速分布散点图

将浮标风速和 SAR 风速分别与浮标测得的气温、海温、气压、有效波高和有效波周期，以及 MERRA-2 的比湿数据输入 COARE V3.5 计算得到不同的海气通量。摩擦风速、风应力、拖曳系数潜热通量和感热通量分别用 SAR 风速数据和浮标风速计算得到，如图 3 所示。摩擦风速的 bias 为 -0.03 m/s，RMSE 为 0.09 m/s，STD 为 0.09 m/s，在 0.2 m/s 以上的风速区间，SAR 风速得到的结果与浮标风速得到的结果一致性较差，且大部分散点在对角线下方，SAR 风速数据得到的结果整体数值要略低于浮标数据。风应力的 bias 为 -0.03 N/m$^2$，RMSE 为 0.09 N/m$^2$，STD 为 0.08 N/m$^2$，散点分布于摩擦风速的分布相似，同样是在数值较高时一致性较差且 SAR 数据获得的值偏低。拖曳系数的 bias 为 -0.29，RMSE 为 0.38，STD 为 0.25，SAR 数据和浮标数据得到的拖曳系数在低值区和高值区都有较明显的不一致性，SAR 风速得到的拖曳系数值低于浮标风速得到的拖曳系数的现象比摩擦风速和风应力更为明显。潜热通量的 bias 为 -8.32 W/m$^2$，RMSE 为 30.72 W/m$^2$，STD 为 29.57 W/m$^2$，感热通量的 bias 为 0.05 W/m$^2$，RMSE 为 7.23 W/m$^2$，STD 为 7.23 W/m$^2$，相较于潜热通量，利用 SAR 得到的感热通量与浮标数据得到的值一致性明显更优，这可能与 COARE V3.5 在计算时未考虑降水带来的影响有关。

将上述结果绘制成随 10 m 高风速的分布散点图，结果如图 4 所示。从图中可看出，摩擦风速和风应力都随 10 m 高风速的增大而逐渐增大，其中摩擦风速与 10 m 高风速近似呈线性分布，在 10 m/s 以上的风速区间，SAR 数据和浮标数据得到的摩擦风速和风应力结果不一致性更加明显。拖曳系数在低风速区间随着 10 m 高风速的增大而逐渐减小，在 2.5 m/s 处达到最低值而后又逐渐增大。潜热通量和感热通量随着 10 m 高风速的增大，散点分布也越分散，但热通量的数值大小也与时间有关。与图 3 的结果相一致，摩擦风速、风应力和潜热通量都是在数值较高时，与浮标计算得到的真实值一致性较差，而拖曳系数和感热通量则是在低值区和高值区都出现了明显的偏差。

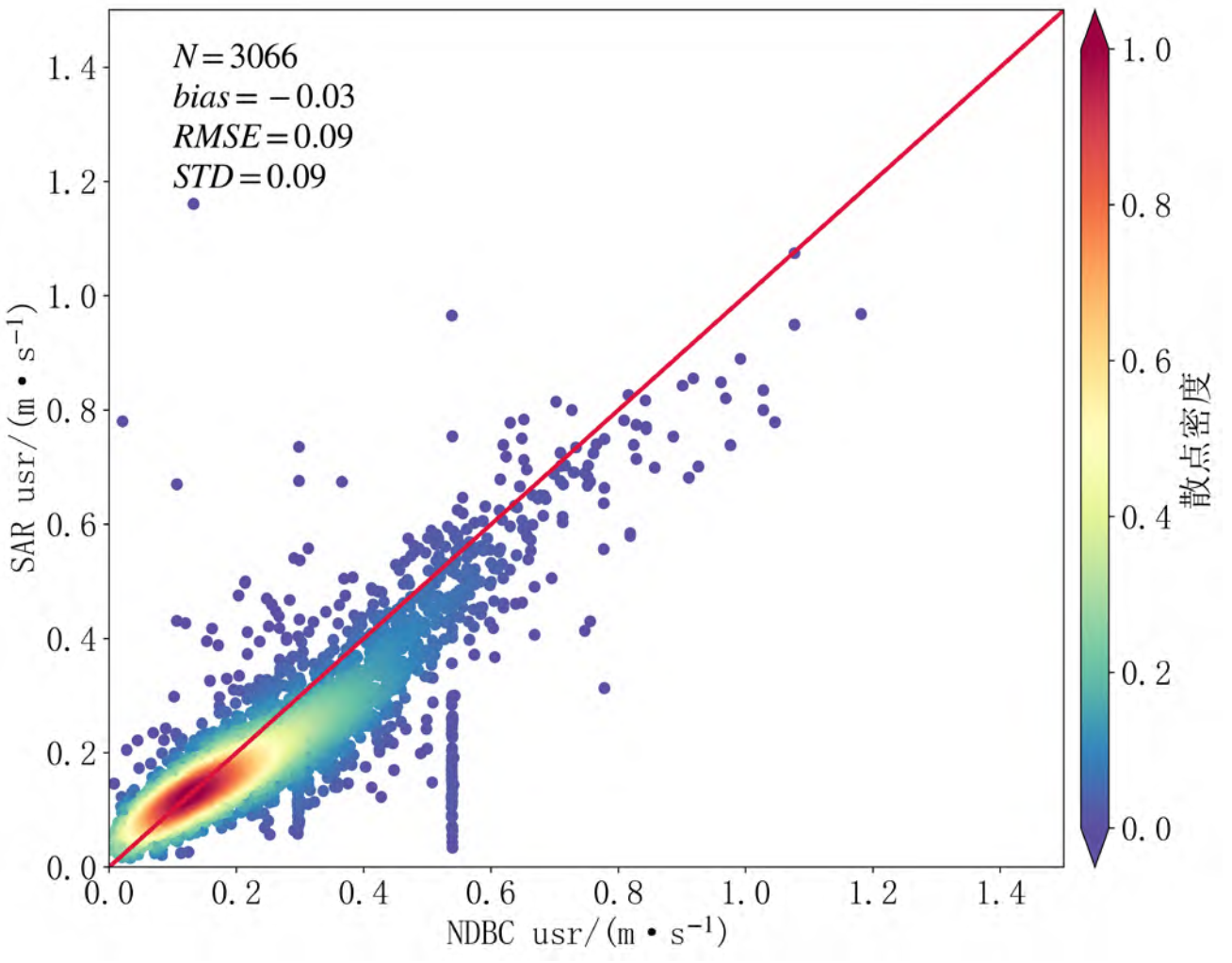

（a）摩擦风速

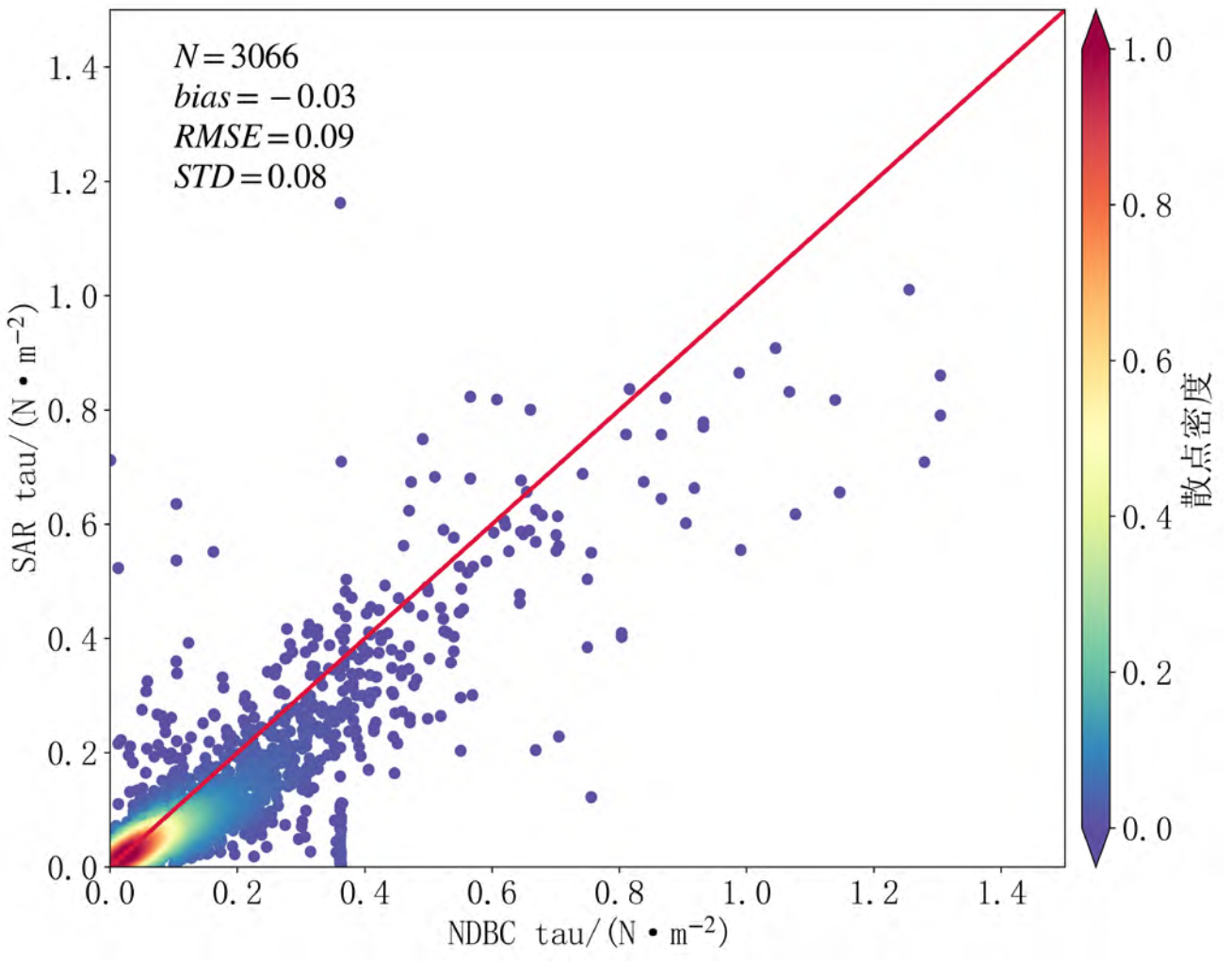

（b）风应力

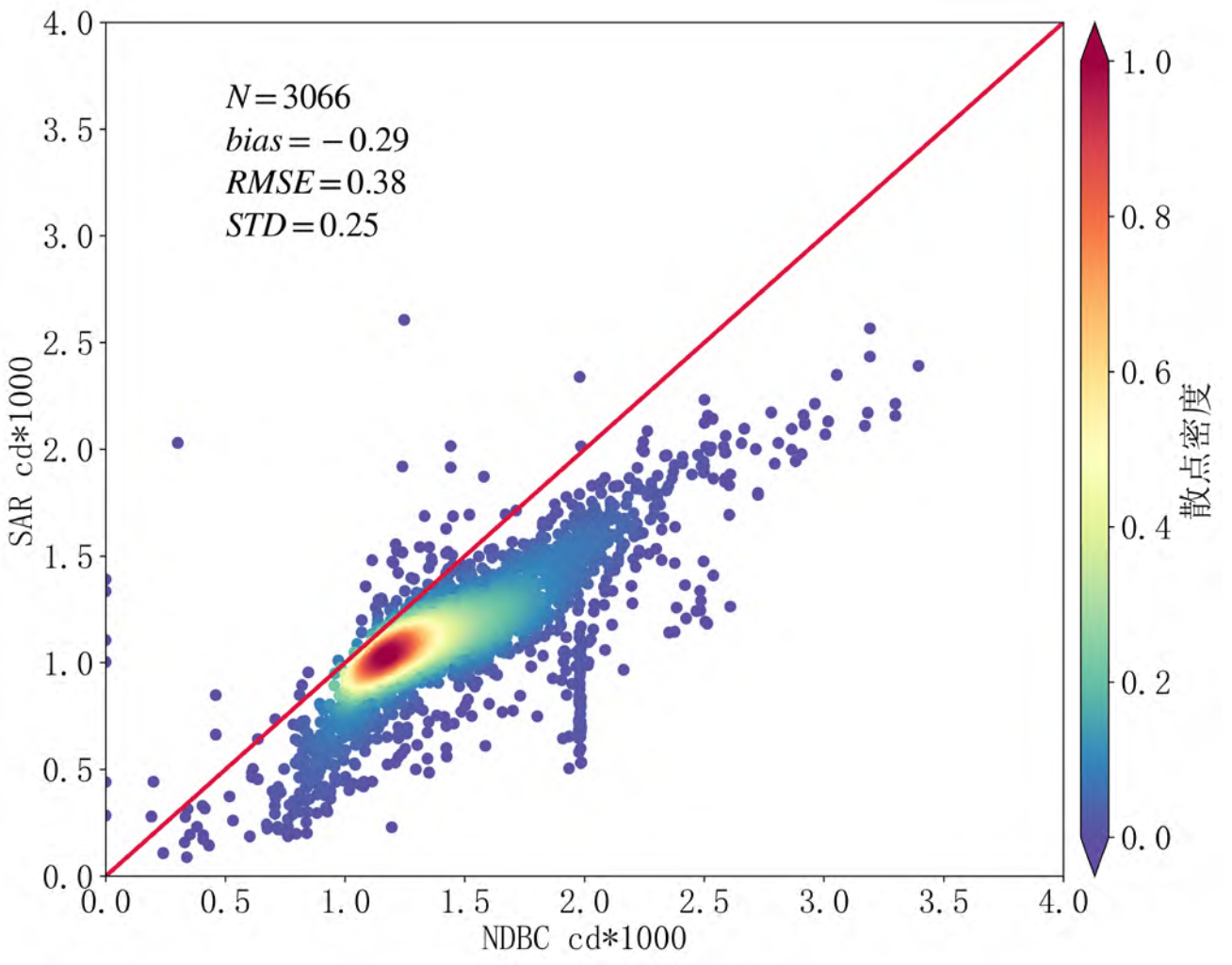

（c）拖曳系数

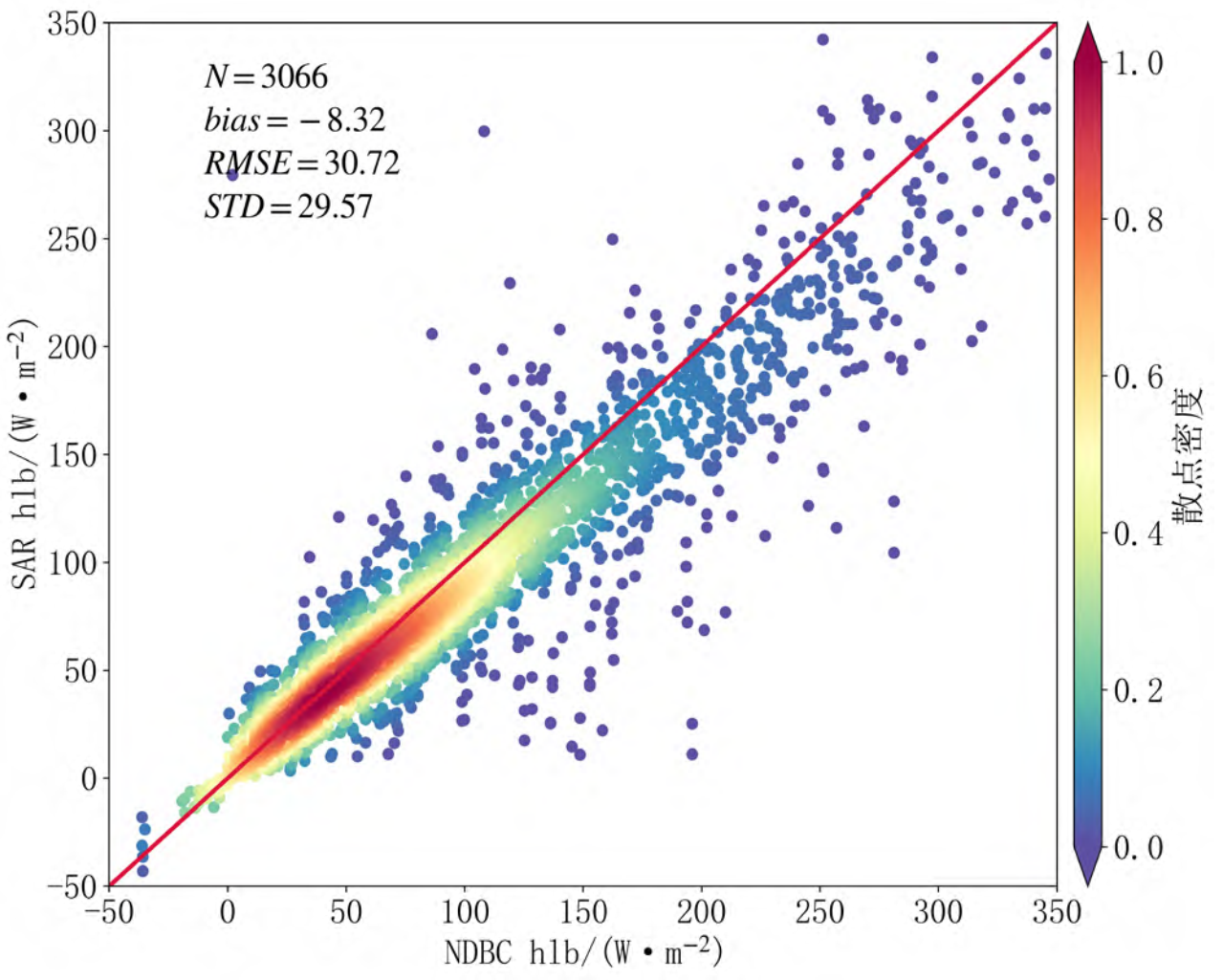

（d）潜热通量

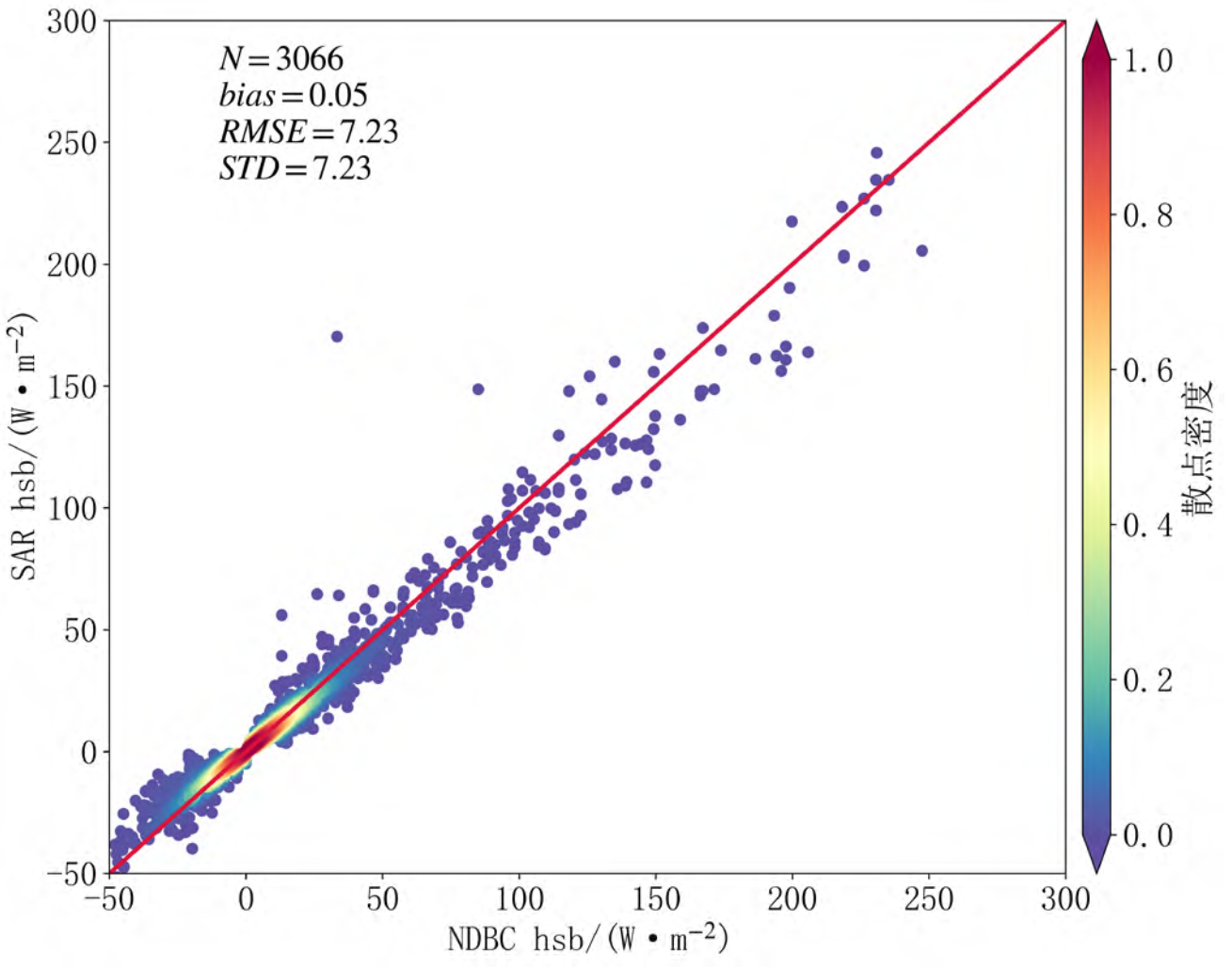

（e）感热通量

图3　用SAR风速数据和浮标风速计算得到的摩擦风速、风应力、拖曳系数、潜热通量和感热通量等物理量的散点图

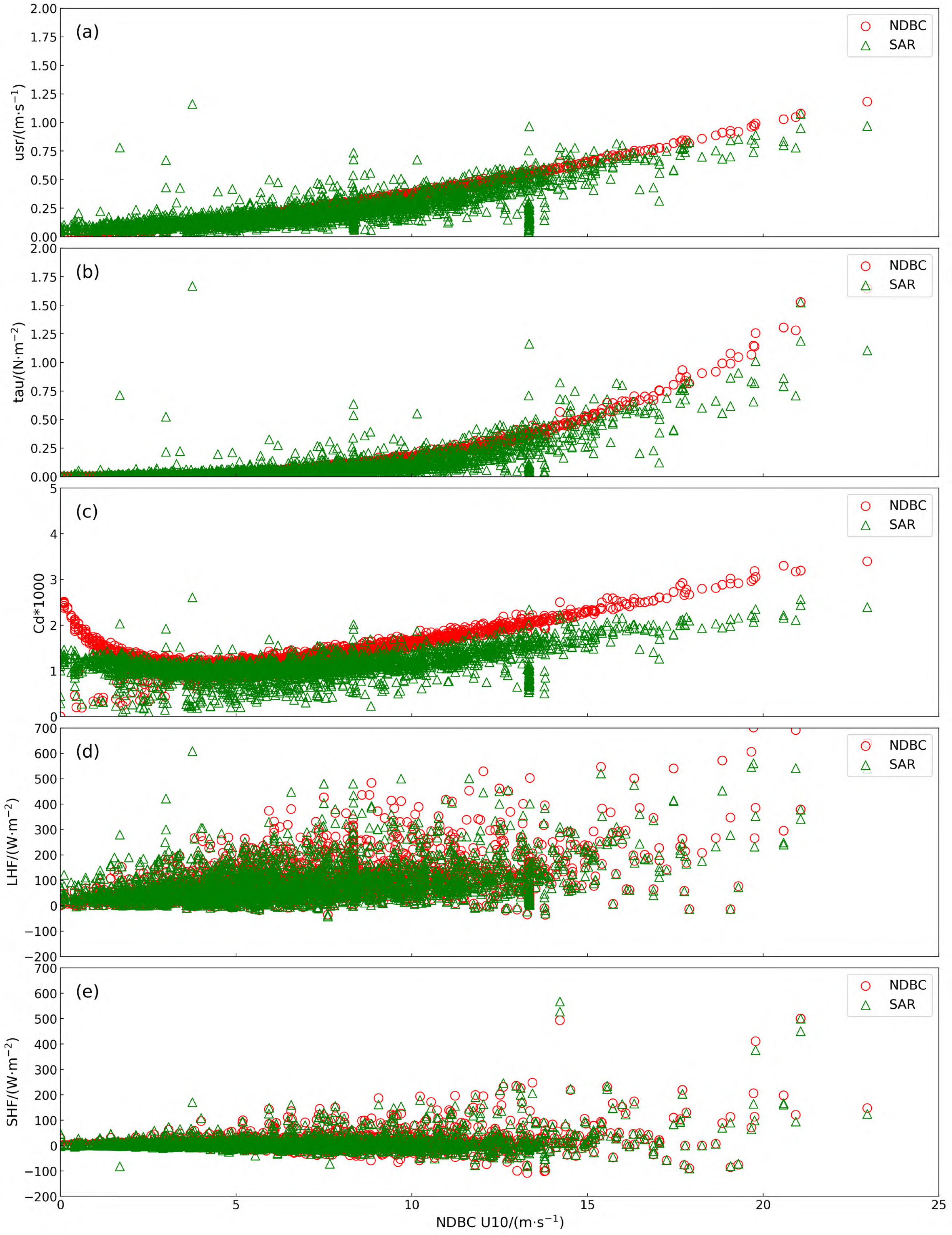

图4　各物理量随海面十米高风速的分布，其中（a）摩擦风速；（b）风应力；（c）拖曳系数；（d）潜热通量；（e）感热通量

## 2.2 神经网络校正SAR风速

用训练集的 SAR 风速数据对 BP 神经网络进行训练，以 NDBC 风速作为真实值，用得到的模型对测试集的 SAR 数据进行校正，得到的 SAR 风速校正结果与浮标实测风速对比如图 5 所示。能看出经过校正后，SAR 风速与浮标风速的偏差明显降低，bias 从 -0.95 m/s 降低至 0.14 m/s，RMSE 从 2.13 m/s 降低至 1.89 m/s，STD 从 1.91 m/s 降低至 1.88 m/s，SAR 风速低估现象得到了明显改善，校正后的 SAR 风速更加接近浮标实测的真实值。

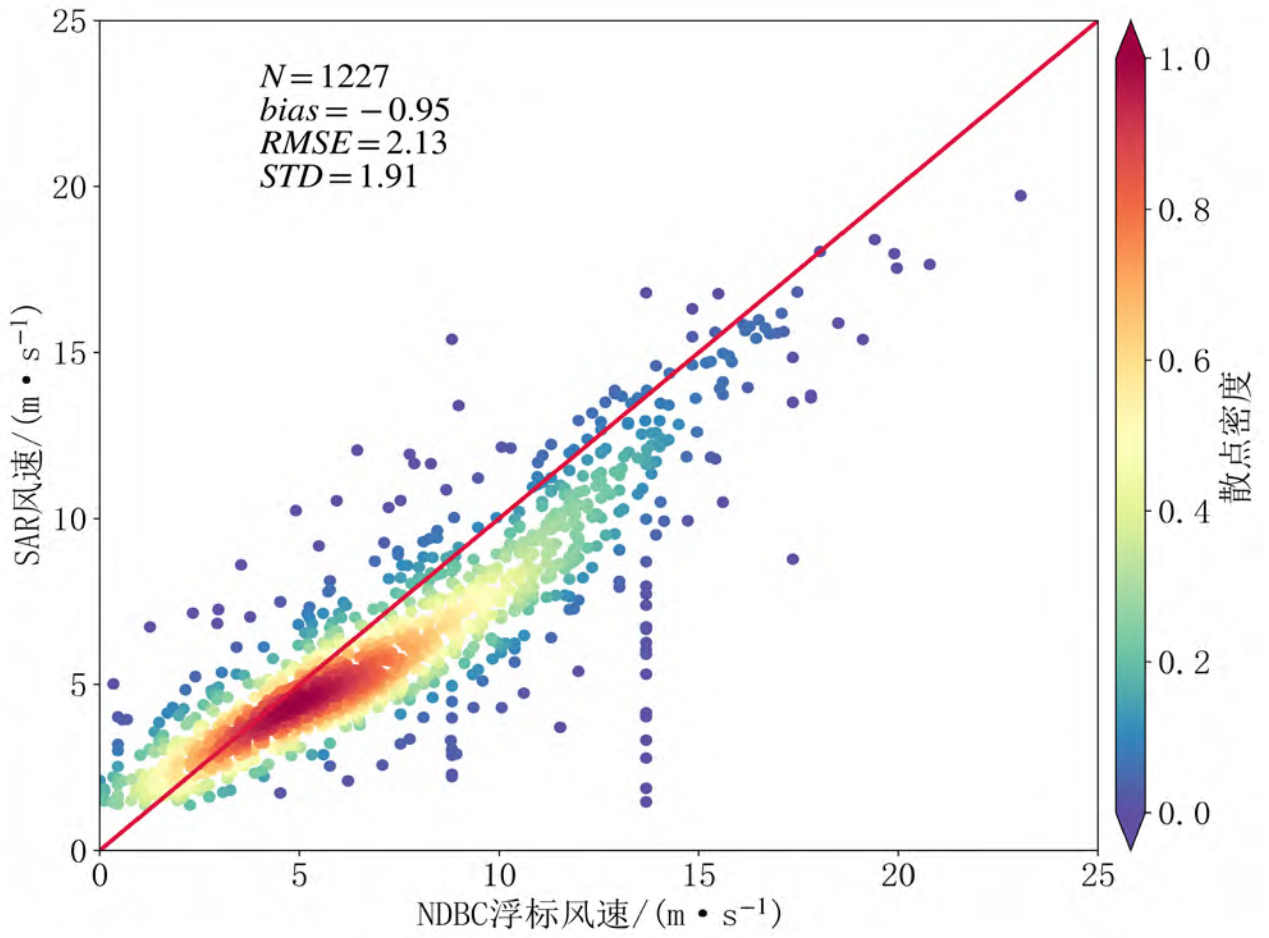

（a）校正前

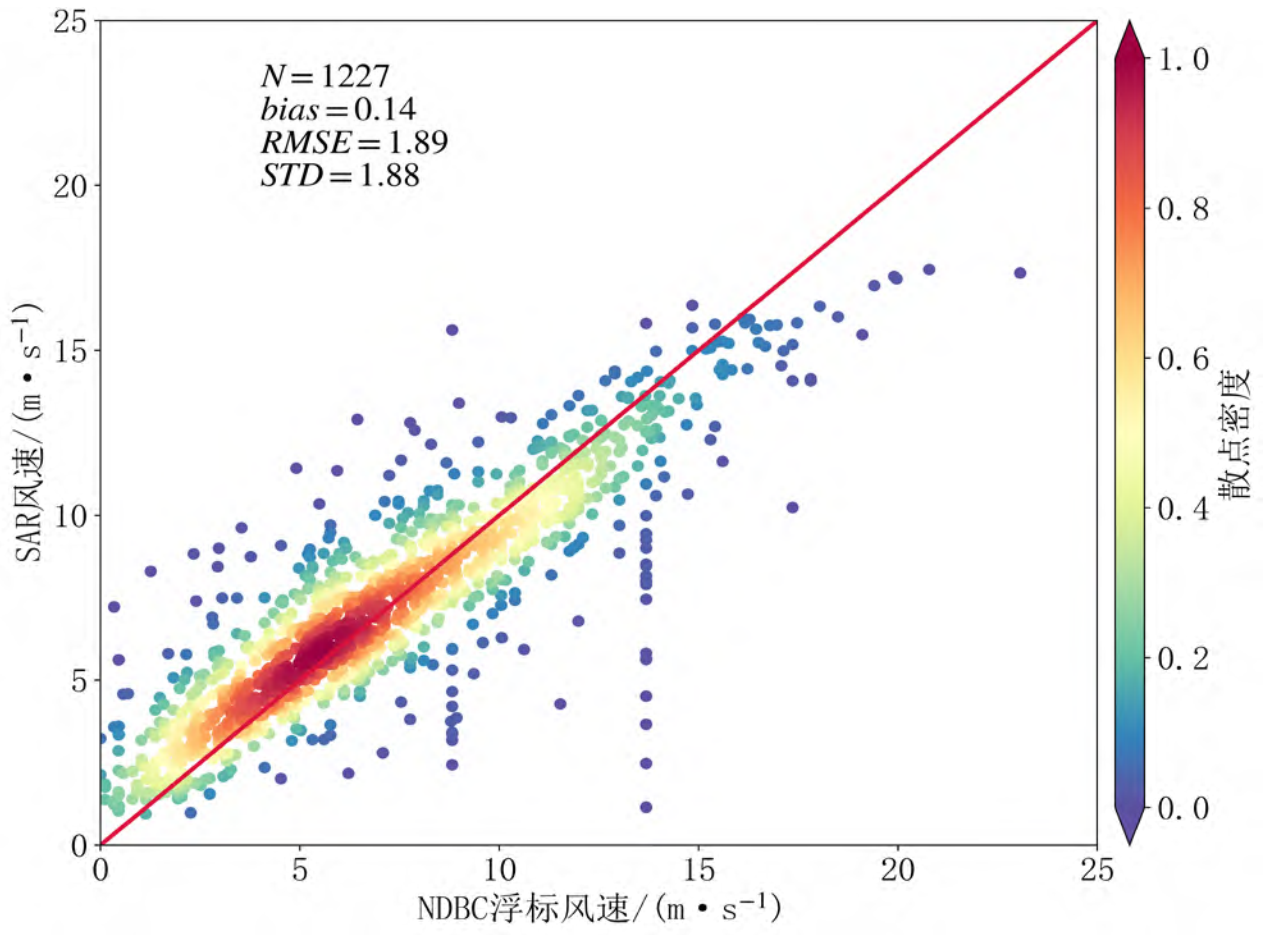

（b）校正后

图5　风速校正前后散点图

将校正后的 SAR 风速与其他物理量输入 COARE V3.5，将得到的结果与用测试集浮标风速计算得到的结果进行比较，图 6 为五组校正前后的密度散点图。通过对比可发现，校正后 SAR 风速计算得到的各海气通量值都更加接近浮标风速计算得到的结果，尤其是 bias 偏差统计量，校正后有了明显的降低。摩擦风速的 bias 从 -0.03 m/s 降低到 0.01 m/s，风应力的 bias 从 -0.03 N/m$^2$ 降低到 0.00 N/m$^2$，拖曳系数的 bias 从 -0.29 降低到 -0.21，且整体低估的情况略有改善，潜热通量的 bias 从 -7.76 W/m$^2$ 降低到 5.41 W/m$^2$，但校正后的散点在 0-100 W/m$^2$ 区间内集中分布在了对角线上方，略微出现高估现象，感热通量校正前后 bias 从 0.67 W/m$^2$ 减小到 0.06 W/m$^2$，但由于校正前 SAR 风速的感热通量计算结果与浮标风速得到的结果一致性已经较好，校正后的散点图改善并不明显。

可以看出用校正后风速得到的摩擦风速和潜热通量虽然误差统计量整体都有所降低，但在数值较低区间校正后反而出现高估现象，这是由于风速校正时，5 m/s 处的低风速区间的风速在校正后出现了高估现象，最终导致了摩擦风速和潜热通量在低数值区也出现了高估，这同时也表明摩擦风速和潜热通量与十米高风速有较为紧密的相关性。

## 3　结束语

本文从影响海上通信的海洋环境要素出发，探究了用卫星遥感资料来研究影响海上无线通讯的海洋环境的可行性。基于哨兵一号卫星 SAR 风速数据，利用海气通量算法 COARE V3.5 计算了海气通量，包括摩擦风速、风应力、拖曳系数、潜热通量和感热通量，这些物理量都是影响海上通信重要环境要素。对比 SAR 风速和浮标实测风速得到的海气界面通量发现，SAR 风速得到的结果虽整体上与浮标风速结果呈线性关系，但在高风速区间内依旧存在比较明显的偏差。用 BP 神经网络对 SAR 风速数据进行校正，使其更加接近浮标风速，用经过校正的 SAR 风速计算海气界面通量，发现结果与浮标风速得到的结果的一致性有所提高。本文结果表明经过神经网络校正后的 SAR 风速资料可以较好地用于研究海洋环境，为海上通信研究提供较为可靠的环境信息，提高海上通信质量。

## 参考文献：

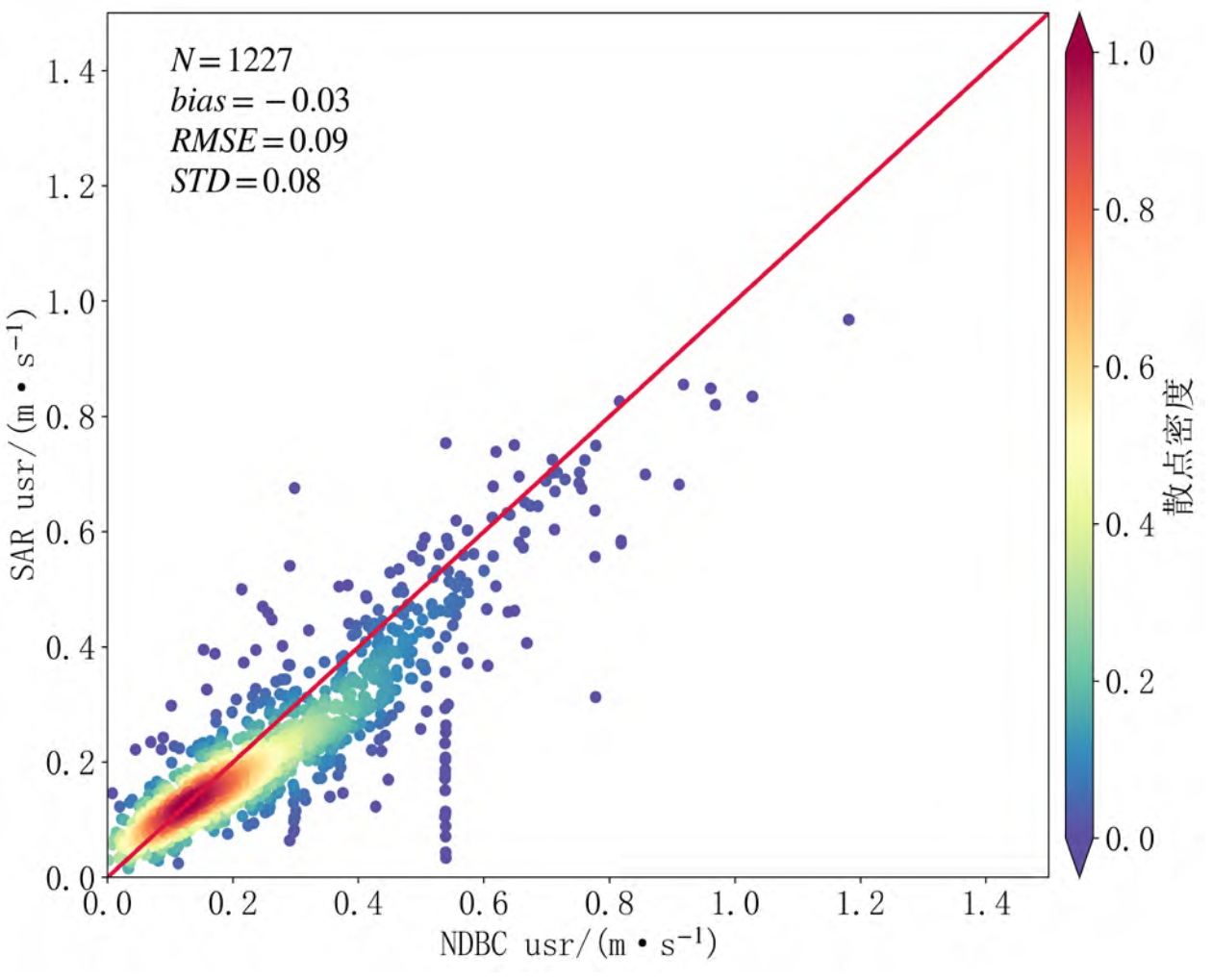

（a）摩擦风速（校正前）

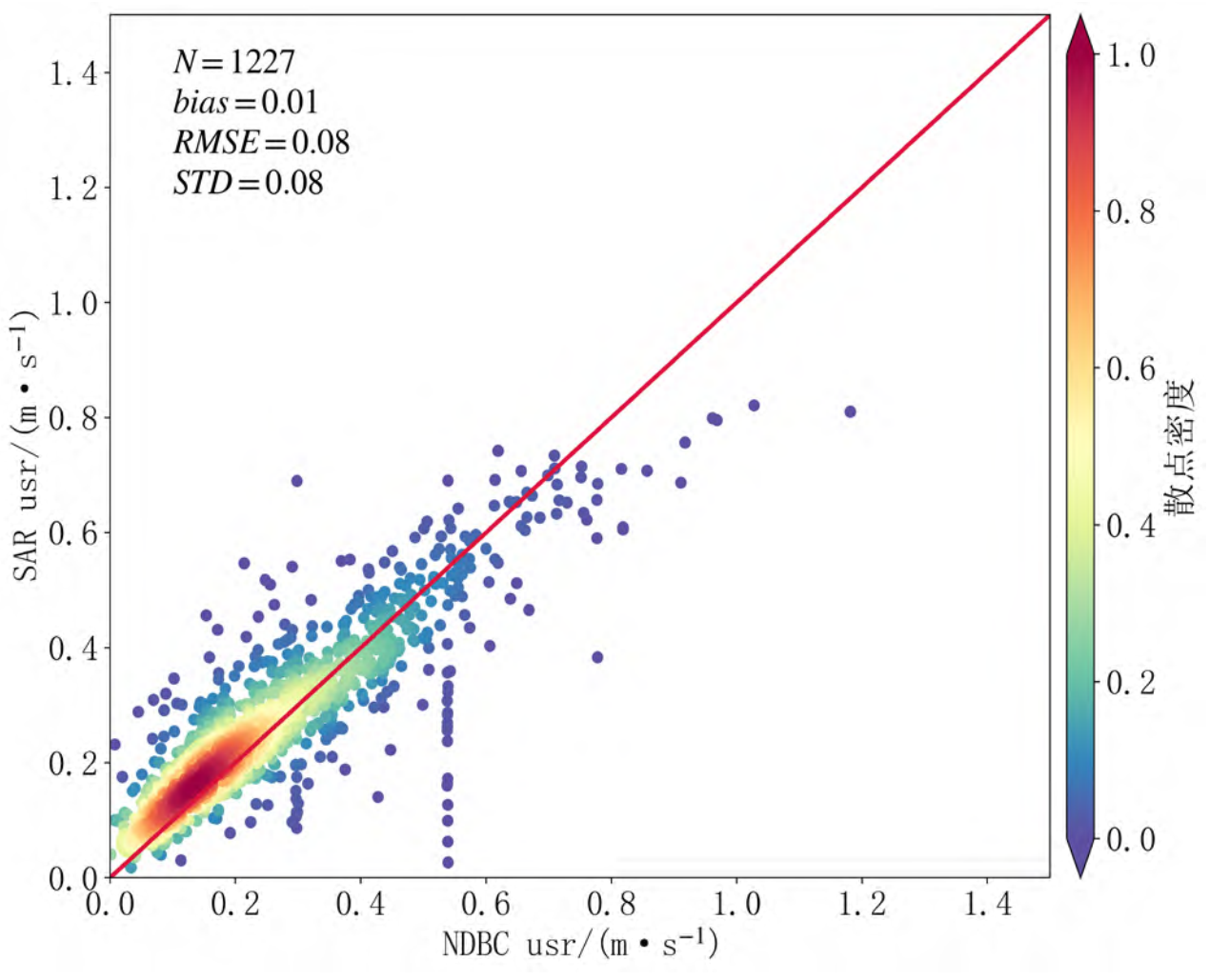

（b）摩擦风速（校正后）

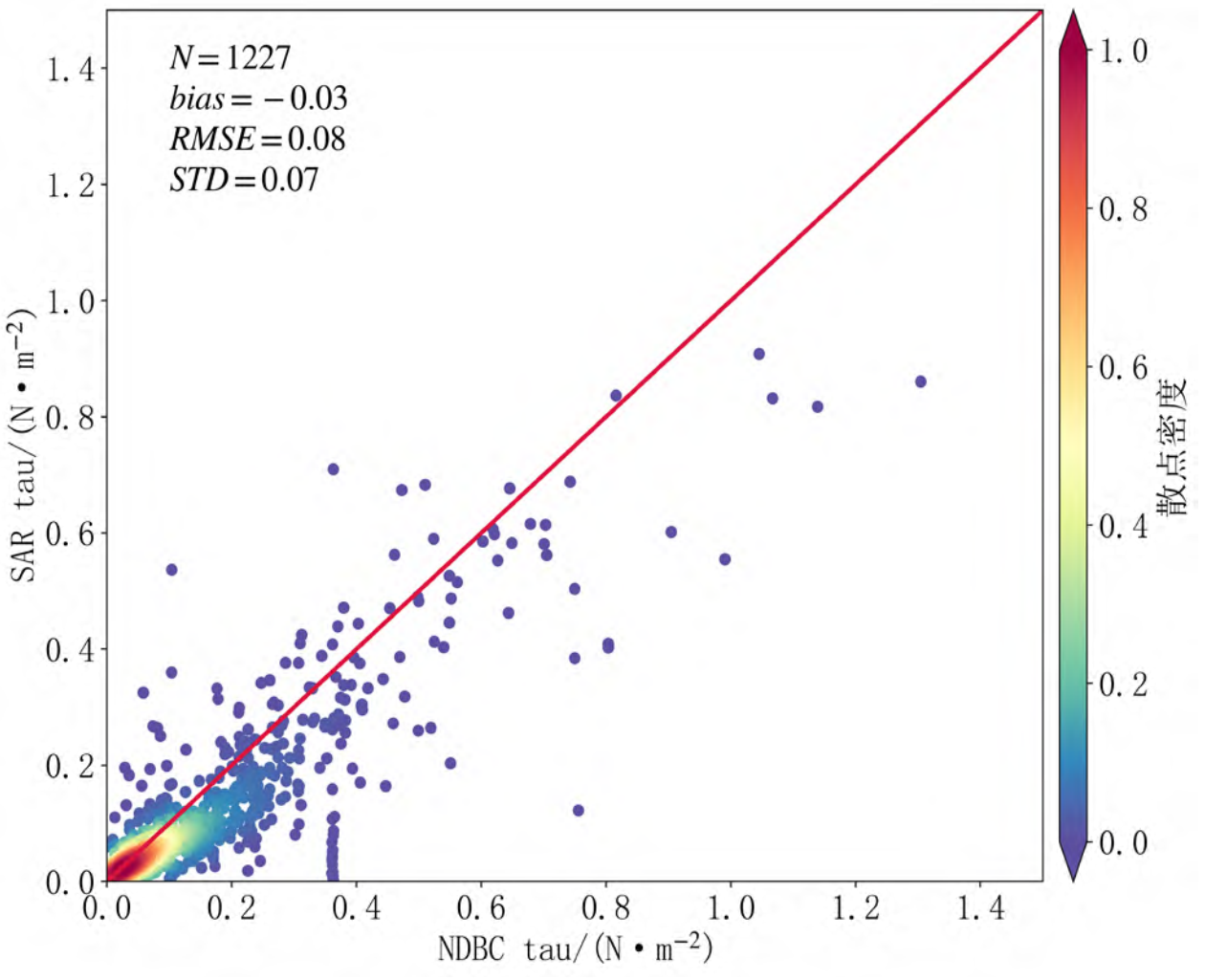

（c）风应力（校正前）

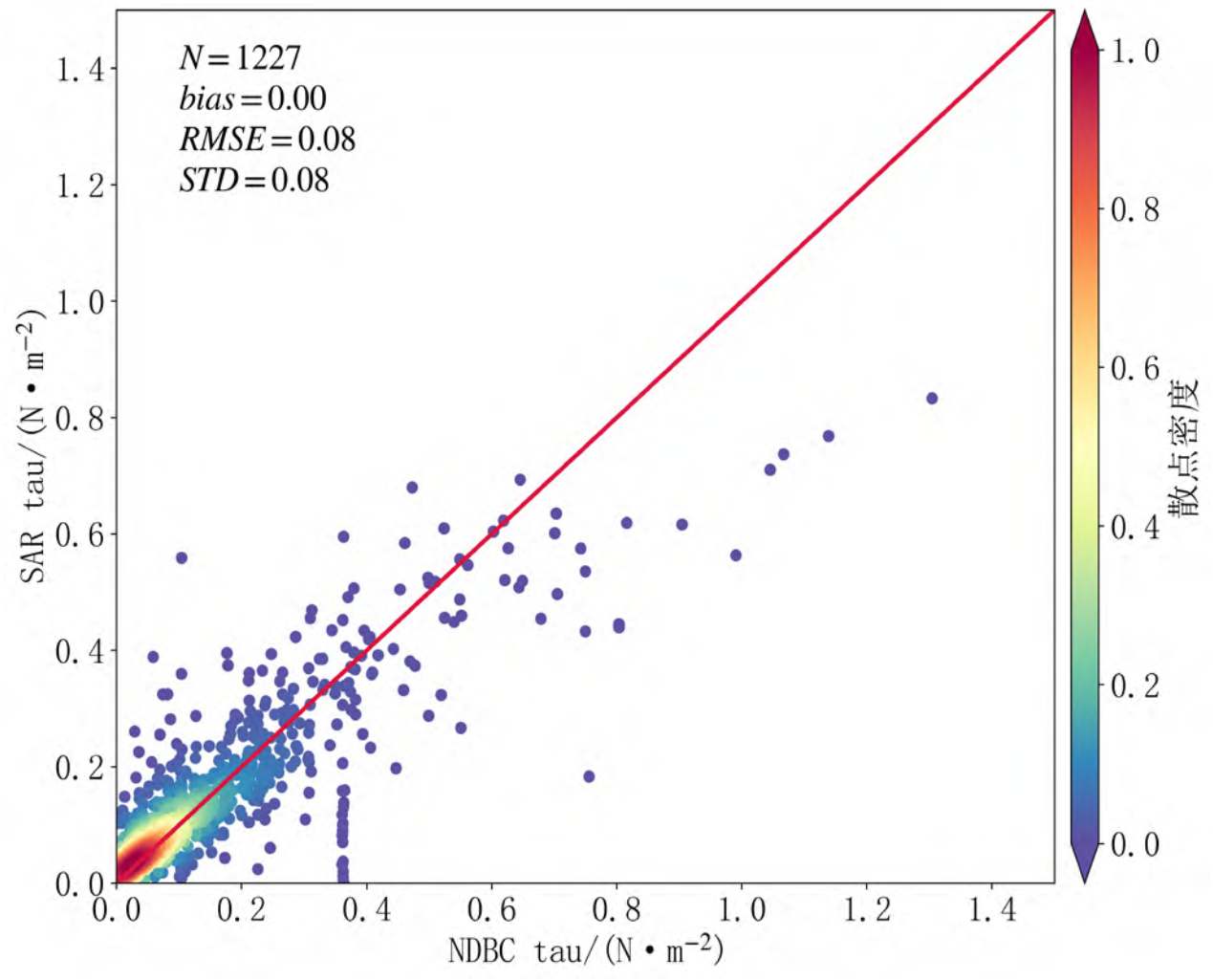

（d）风应力（校正后）

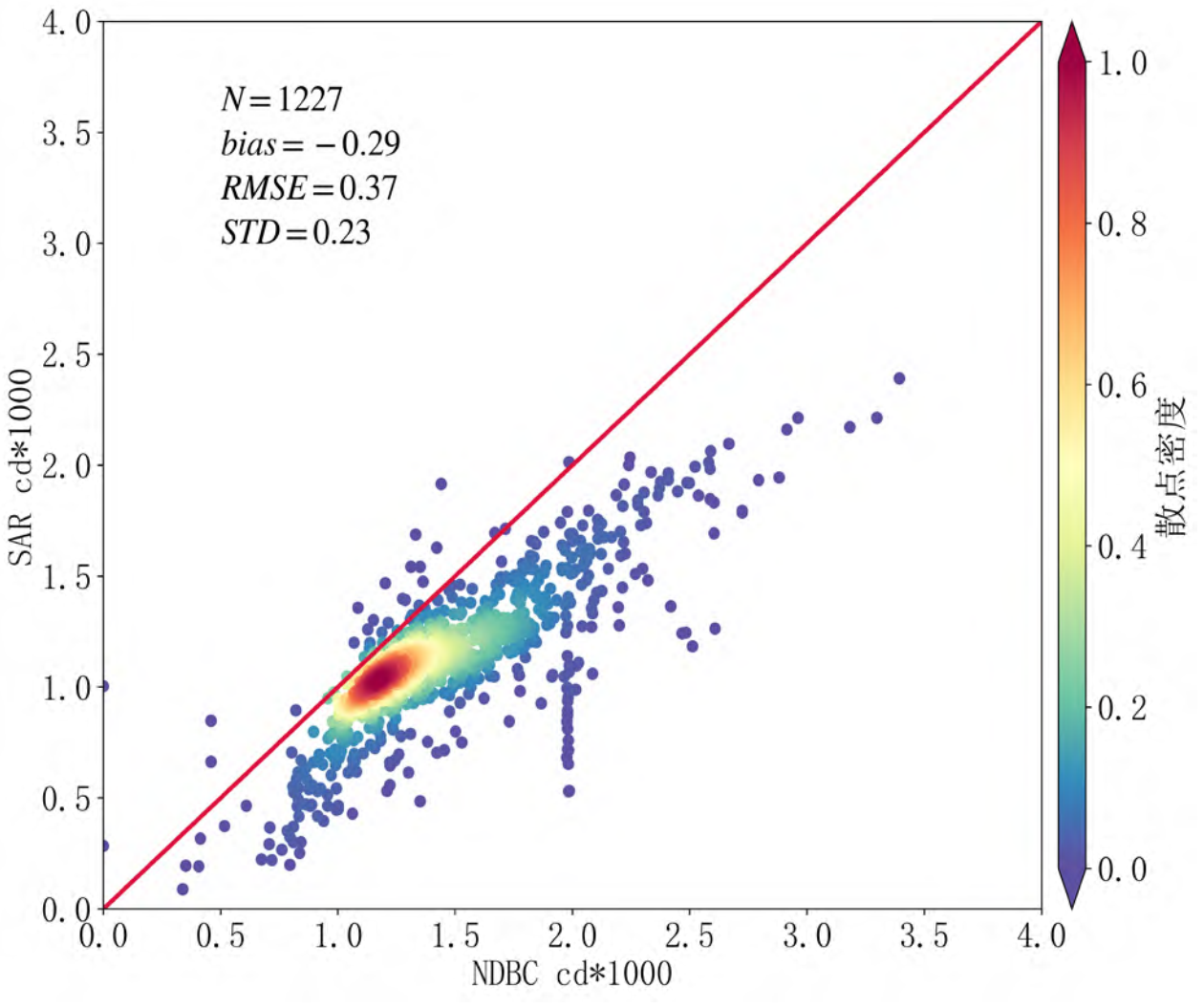

（e）拖曳系数（校正前）

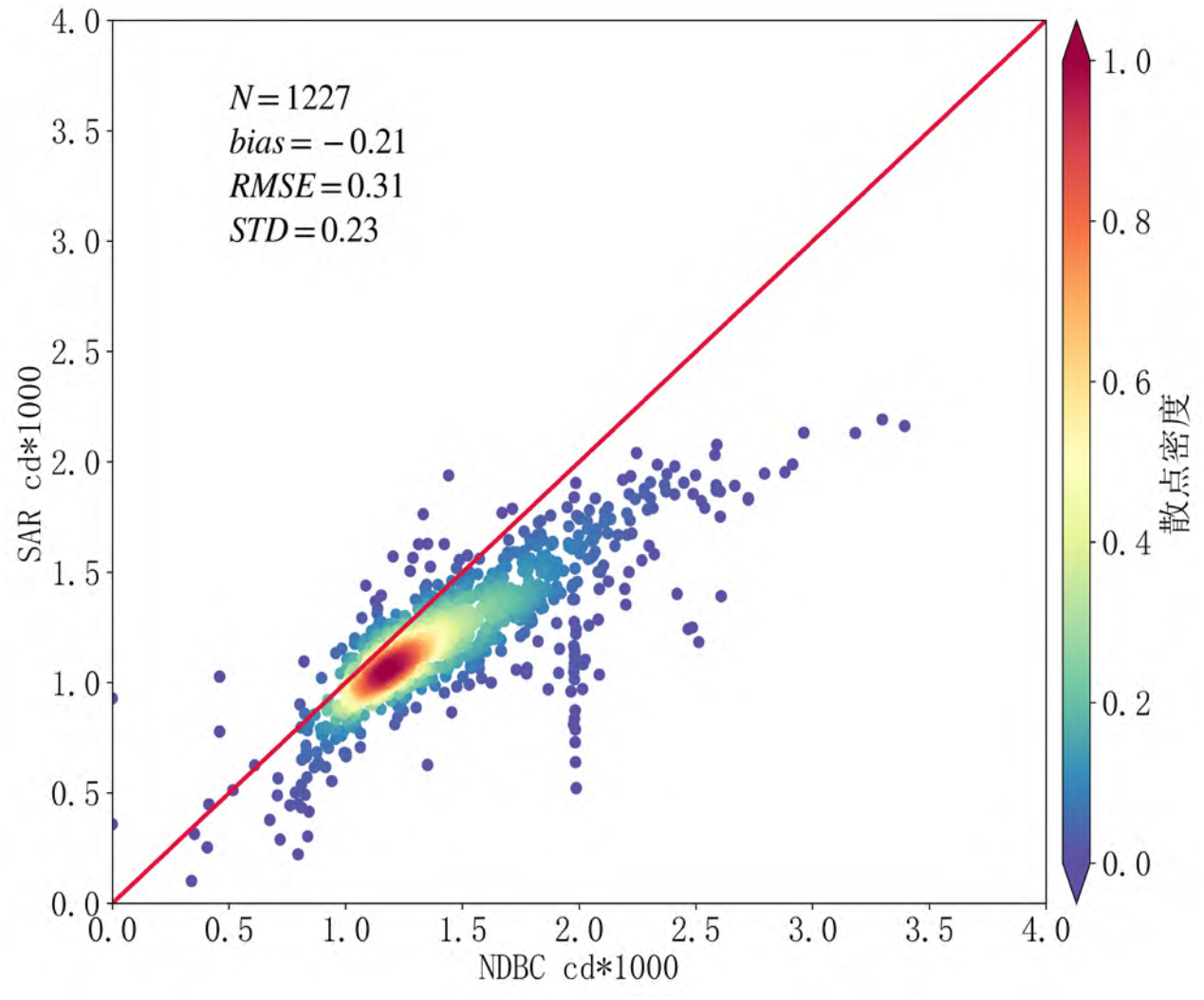

（f）拖曳系数（校正后）

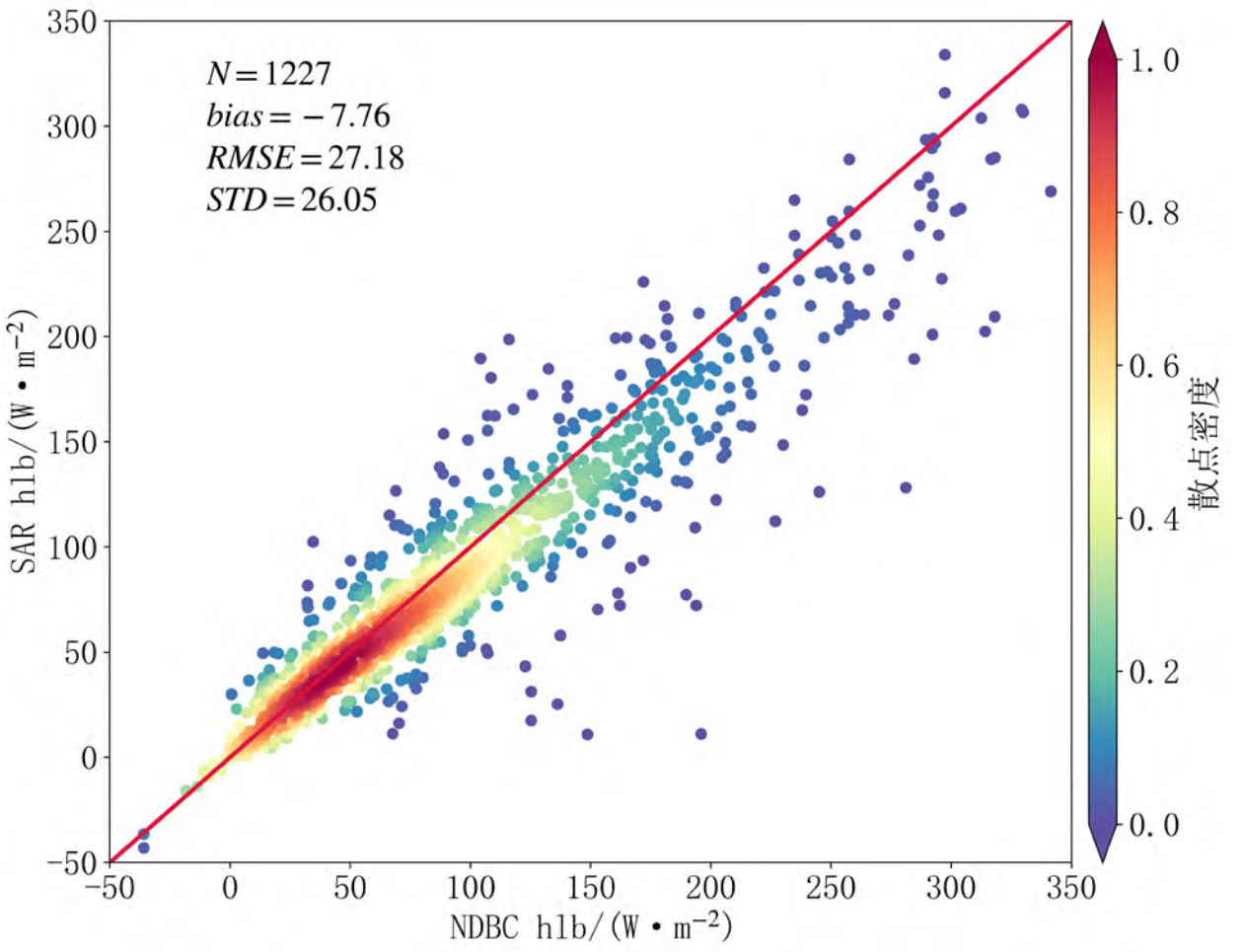

（g）潜热通量（校正前）

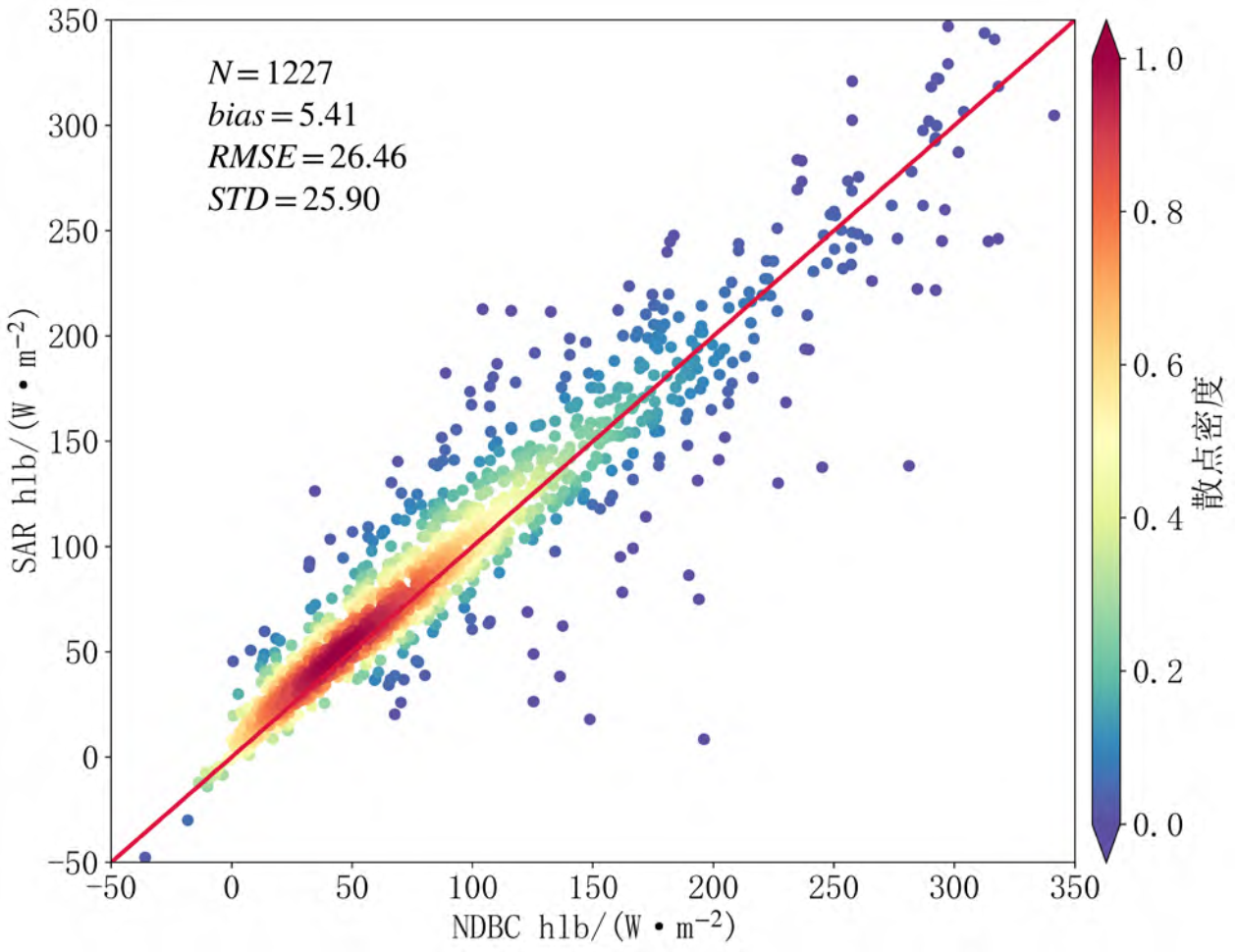

（h）潜热通量（校正后）

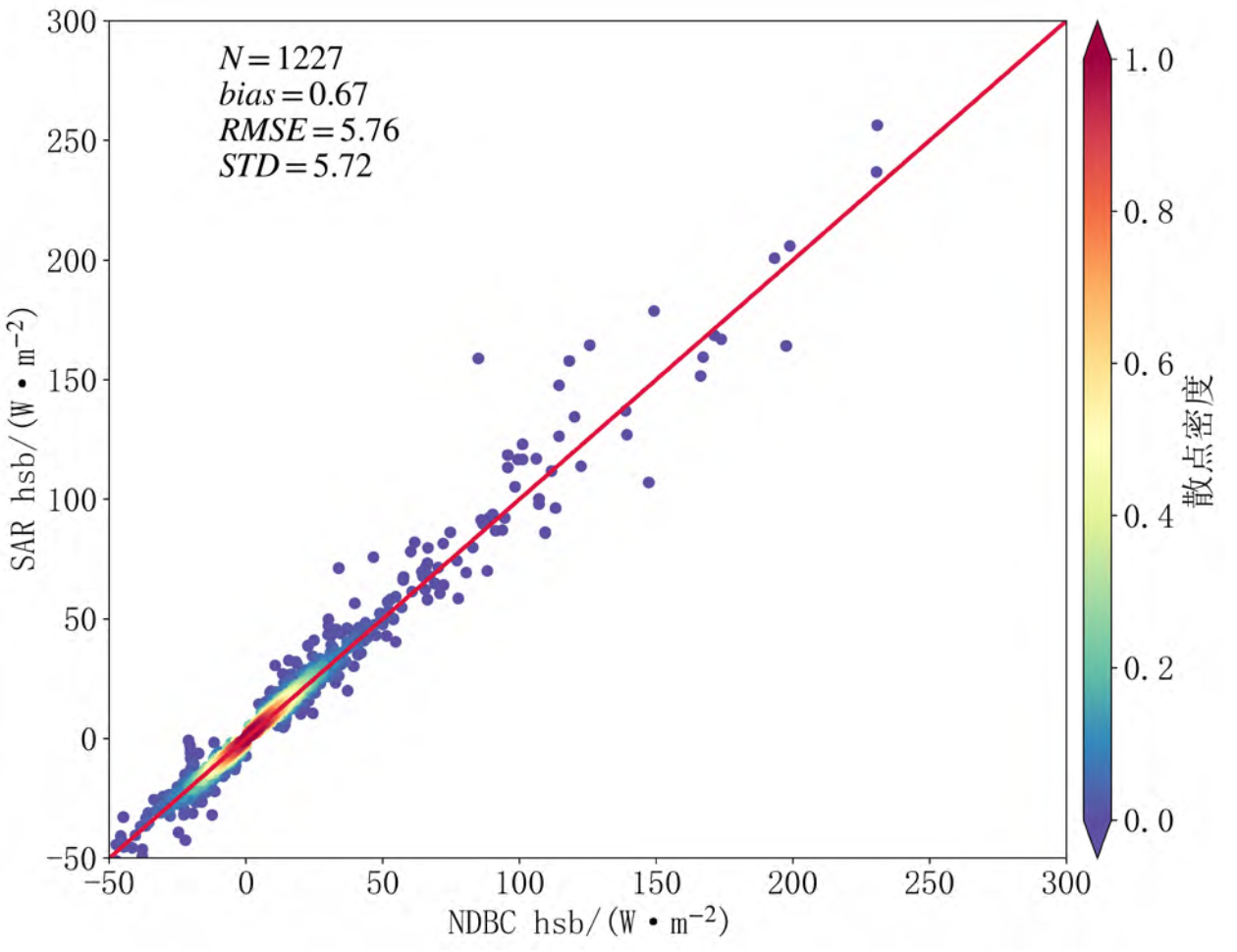

（i）感热通量（校正前）

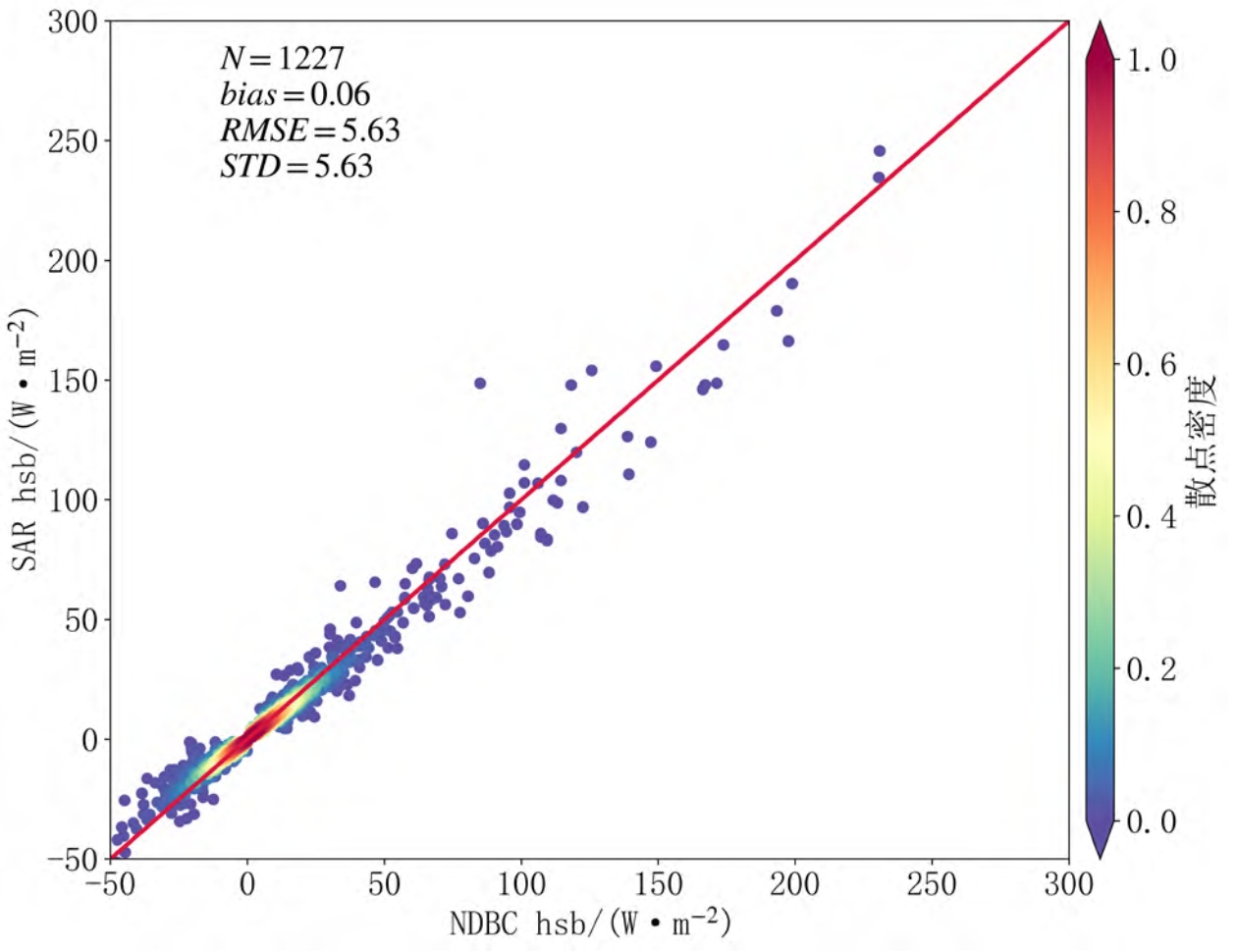

（j）感热通量（校正后）

图 6　BP 神经网络校正前后结果对比

（下转第85页）

## 作者简介

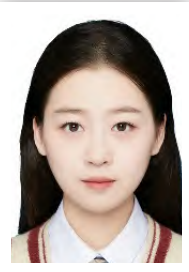

吴佳丽：北京科技大学在读本科生，主要从事6G移动通信、数字孪生技术的研究工作。

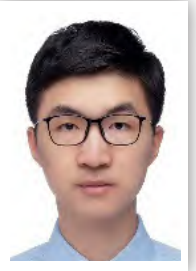

刘向南：博士毕业于北京科技大学，主要从事智能通信、空天地一体化网络的研究工作。

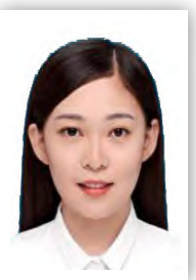

孙春蕾：现任北京科技大学讲师，主要从事通信感知一体化、车联网等技术研究工作。

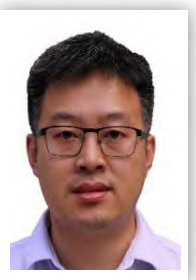

张海君：现任北京科技大学教授、博士生导师，主要从事6G移动通信、人工智能与无线网络的研究工作。

（上接第44页）

## 作者简介

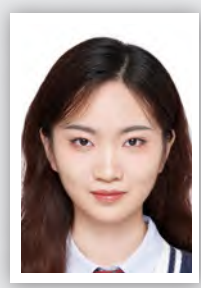

倪晗玥：上海交通大学在读硕士研究生，主要方向为海洋微波遥感、海气界面通量参数化等。

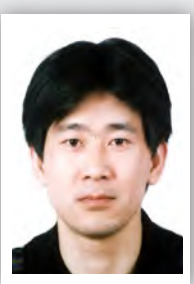

杨劲松：研究员、博士生导师，博士毕业于青岛海洋大学，现任职于自然资源部第二海洋研究所，主要从事海洋微波遥感与卫星海洋学研究。

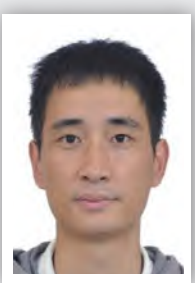

任林：副研究员、硕士生导师，博士毕业于南京理工大学，现任职于自然资源部第二海洋研究所，主要从事卫星海洋微波遥感研究。